\documentclass[fleqn,usenatbib]{mnras}

\usepackage{newtxtext,newtxmath}
\usepackage{graphics}
\usepackage{epsfig}	
\usepackage{amsmath}	

\usepackage{amssymb}	
\usepackage[T1]{fontenc}

\DeclareRobustCommand{\VAN}[3]{#2}
\let\VANthebibliography\thebibliography
\def\thebibliography{\DeclareRobustCommand{\VAN}[3]{##3}\VANthebibliography}

\title[Searching for anomalies]{Searching for short-timescale radio anomalies using nonlinear dimensionality reduction techniques}

\author[X. Yang et al.]{
X. Yang$^{1,2,3}$\thanks{E-mail: yangxuan@pmo.ac.cn, (XY); George.Hobbs@csiro.au, (GH); sbzhang@pmo.ac.cn, (SBZ); xfwu@pmo.ac.cn, (XFW)}, G. Hobbs$^{3}$, S.-B. Zhang$^{1,3}$, A. Zic$^{3}$, L. Toomey$^{3}$, Y. Li$^{2}$, J.-S. Wang$^{4}$,
\newauthor S. Dai$^{3}$, X.-F. Wu$^{1,2}$ 
\\
$^{1}$ Purple Mountain Observatory, Chinese Academy of Sciences, Nanjing 210023, China\\
$^{2}$ School of Astronomy and Space Sciences, University of Science and Technology of China, Hefei 230026, China\\
$^{3}$ CSIRO Space and Astronomy, Australia Telescope National Facility, PO Box 76, Epping, NSW 1710, Australia\\
$^{4}$ Max Planck Institute for Plasma Physics, Boltzmannstra{\ss}e 2, D-85748 Garching, Germany\\
}

\pubyear{2024}

\begin{document}
\label{firstpage}
\pagerange{\pageref{firstpage}--\pageref{lastpage}}
\maketitle

\begin{abstract}
We have searched for anomalous events using  2,520 hours of archival observations from Murriyang, CSIRO’s Parkes radio telescope. These observations were originally undertaken to search for pulsars. We used a machine-learning algorithm based on ResNet and Uniform Manifold Approximation and Projection (UMAP) in order to identify parts of the data stream that potentially contain anomalous signals. Many of these anomalous events are radio frequency interference, which were subsequently filtered using multibeam information. We detected 202 anomalous events and provide their positions and event times. 
Our results show that the UMAP unsupervised machine learning pipeline effectively identifies anomalous signals in high-time-resolution datasets, highlighting its potential for use in future surveys. However, the pipeline is not applicable for standard searches for dispersed single pulses. We classify the detected events and, in particular, we are currently unable to determine the possible origin of events that last multiple seconds. For these we encourage follow-up observations. 
 
\end{abstract}



\section{Introduction}

Radio astronomy is a subject full of surprises. From 1933 to 1997, 17 key discoveries in radio astronomy were found~\citep{Wilkinson2004}. These discoveries can be divided into two classes, seven were made by
testing a theory prediction (these were the ``known–unknowns''), and ten were unexpected results found by chance (the ``unknown–unknowns'')~\citep{Norris2017}. One unexpected highlight in the 21st century was the discovery of Fast Radio Bursts (FRBs). FRBs are bright radio pulses with a short duration (typically milliseconds) and the first FRB was discovered in archival data from CSIRO's Murriyang Parkes 64 m-diameter radio telescope~\citep{Lorimer:2007qn}. It took six more years before \cite{FRB2013} found four more FRBs. 
The large implied distances and the extremely high energies of FRBs mean they are intrinsically interesting and can be used for cosmological studies. As FRBs can now be considered as ``known-unknown'' sources, the Canadian Hydrogen Intensity Mapping Experiment (CHIME), the Australian SKA Pathfinder~(ASKAP) and many other telescopes explicitly design their pipelines in order to find more FRBs and the number of known sources is rapidly increasing.

Most of the survey-related software in radio astronomy has been designed for searching for signals with specific characteristics. For example, \emph{\sc PRESTO}\footnote{\url{https://www.cv.nrao.edu/~sransom/presto/}}~\citep[]{presto} can search for pulsar and FRB signals which are characterised with a dispersion measure~(DM) and pulse width and, for pulsar searching, the pulsation period. \emph{\sc HEIMDALL}\footnote{\url{https://sourceforge.net/projects/heimdall-astro/}} has been explicitly designed to search for FRBs following the dispersion measure law. As quantified by~\cite{suki}, these software packages become less effective as the signal deviates further from the expected properties.

We know that the radio data sets likely contain currently undetected ``unknown-unknown'' sources. For instance, an ultra-long-period~(18.18 min)  radio source GLEAM-X J162759.5-523504.3 has been detected with Murchison Widefield Array~\citep{ulp} and the pulse duration varies from 30 to 60s. Another source, GPM J1839-10, found by \cite{mwa_white_dwarf} also shows a long-term~(21 min) radio periodicity, where each pulse persists for 30 to 300s. 
\cite{54minulp} found ASKAP~J1935+2148 with period of $\sim$ 54 minutes when following-up on gamma-ray burst observations using ASKAP. This particular source has detected pulse widths between 10 to 50 seconds in the ASKAP observations, but with typical pulse widths of 370ms as detected in follow-up observations with the MeerKAT telescope. 

Large-scale surveys are being carried out with radio telescopes in order to search for technosignatures~\citep{seti1,seti2}.
A complete understanding of these radio emissions on various time scales is still not fully achieved.  No convincing detection of extraterrestrial technosignatures has been made and the nature of any such signals is not known.  Hence, such surveys require searching large regions of parameter space with highly flexible algorithms.

One approach for finding the ``unknown-unknowns'' is machine learning. \cite{forest_lightcurve2020} used HDBSCAN and t-SNE as anomaly detection algorithms and found seven uncatalogued variables and two stellar flare events. \cite{proximityclustering2019} utilized the t-SNE algorithm to identify <4\% of each season's data as outlying in the Kepler field data. \cite{ASKAP_Peculiar} trained a self-organizing map~(SOM) to select 0.5\% of the most complex and peculiar sources from ASKAP data.
\cite{umap_grav} combined a deep residual network~(ResNet) and UMAP to search for spectrograms with anomalous patterns in public LIGO data.

Our research focuses on the Parkes archive pulsar search mode data~\citep{george_archive}. 
\cite{songbo_database} found five previously undetected Rotating radio transients~(RRATs) and one FRB in the Parkes archive dataset from 1997 to 2001.
\cite{songboburst2019} found an FRB in the same dataset as the Lorimer burst, a remarkable twelve years after the discovery of the latter. We believe there are still numerous hidden signals of interest waiting to be detected in this archive. 

Previous machine learning algorithms that have been developed to search for anomalies follow a procedure similar to the following:

(1) They first extract features from the raw astronomical data.

(2) Dimensionality reduction for high-dimensional data is carried out to enable the resulting data to be visualised.

(3) The low-dimensional data are clustered, and anomaly ranking or outlier detection methods are applied.

Dimensionality reduction techniques play an important role in visualizing high-dimensional data. One widely used method is Principal Component Analysis (PCA), a well-established linear technique. PCA involves combining the original variables to derive principal components, enabling a reduction in the dimensionality of the data~\citep{pca}.
Nevertheless, nonlinear techniques continue to grow in popularity and are better suited for effectively addressing the issue of overcrowding in data feature extraction.
Prior to 2018, the t-Distributed Stochastic neighbour Embedding (t-SNE) was the most popular nonlinear technique in single-cell analysis~\citep{tsne}. However, UMAP has since become more favoured among scientists due to its swiftness and ability to retain both local and global data structures~\citep{umap0}.

In our paper, we implemented ResNet, UMAP, and spectral clustering to accomplish the three stages. This pipeline was applied to datasets from the Parkes archive to search for any possible outlier. The time-frequency images derived from the raw data were used. In Section~\ref{sec:data}, we provide a description of the data and algorithms used in this study. The results of our clustering analysis are presented in Section~\ref{sec:result}, while Section~\ref{sec:Dis} includes a discussion of our findings.


\section{Methods}
\subsection{Data processing}
\label{sec:data}
The data used in this study were obtained from the Parkes archival data P269, a deep survey of the Large and Small Magellanic Clouds~\citep{P269_1,P269_2}. This particular dataset was selected as a representative sample of archival data containing anomalous events, as it includes the first detected FRB and another weaker FRB. It encompasses the initial discovery of FRB~\citep{Lorimer:2007qn}, and another weaker FRB was detected from this dataset twelve years later~\citep{songboburst2019}.  
The sample time, spectra per subintegration, central frequency, bandwidth, and number of channels for these observations were 1000$\mu s$, 4096, 1374\,MHz, 288 MHz, and 96, respectively. We processed all of the data in the observing semester P269-2001JANT, which contains 2,258 files and 2,520 hours of integration time.

\begin{figure*}
    \begin{center}
    \includegraphics[width=0.95\textwidth]{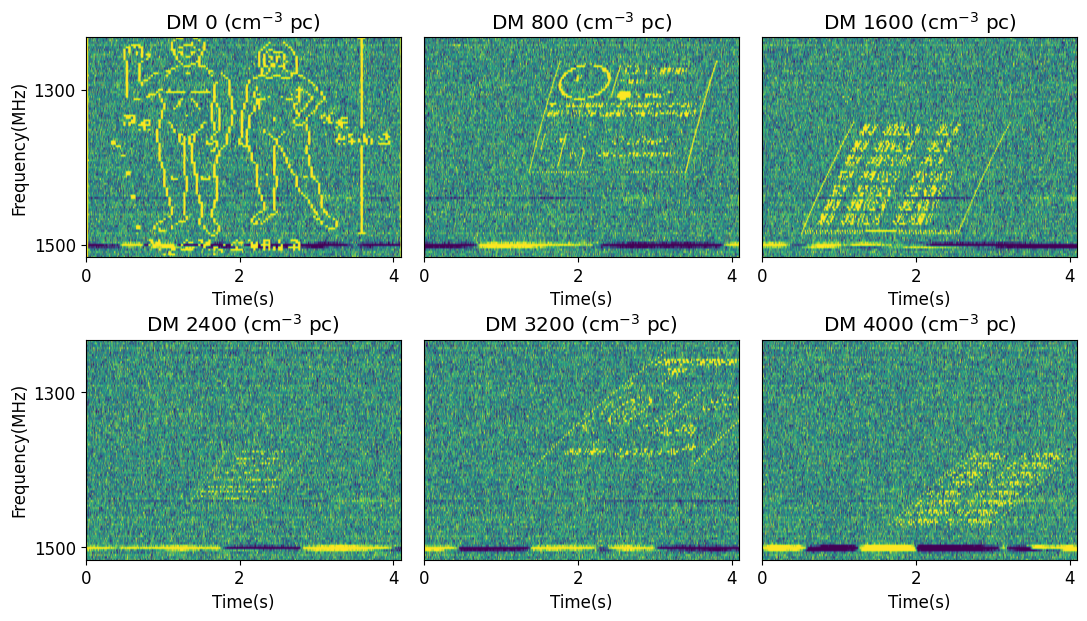}
    \caption{Six samples of the injected Cosmic Call data. Each figure refers to 4.096 seconds of integration time. The title of each figure shows the DM of the injected Cosmic Call data.}
    \label{figure:cosmic}
    \end{center}
\end{figure*}

We also converted some artificial signals into time-frequency pixels, and injected them into the real observation data using the software described in \cite{simulate}. The artificial signals were downloaded from the Cosmic Call 2003 message~\citep{cosmic}. As shown in Fig.~\ref{figure:cosmic}, the 22 Cosmic Call images were resized into four distinct sizes, and subsequently distributed across a range of DM values, spanning from 0 to 4,000 with a step size of 800. This batch of simulation data was later used as a tracer in order to help locate the possible anomalies in the real data. 


In brief, for our processing:
\begin{enumerate}
\item We selected the PSRFITS input files in groups of three. For each group, we also added in the Cosmic Call data file that will be used as a tracer for anomalous events.
\item 
Each of these four data files was first read by the python package \emph{\sc astropy}. Almost all files contained 2,051 subintegrations (shorter files are calibration observations and files which ended early). Each subintegration containing $96\times4096$ points was resized and output as a $256\times256$ image. The image refers to 4.096 seconds of observation. We therefore obtain 2,051 images for each of the four data files.
\item We then employed a pre-trained ResNet50V2 model to extract features from each image and obtained a 512D feature vector that represents a compressed version of the original image. \label{part3}
\item The 512D feature vectors were then input to UMAP. This gives us a UMAP embedding for all the images in the four input files. Similar-looking images will appear close to each other in this UMAP embedding. \label{part4}
\item We searched for different clusters in this UMAP embedding using ``spectral clustering'' and silhouette coefficient values.\label{part5}
\item We now have identified the different clusters corresponding to the images produced from the four input files.  The clusters that contain more than 50\% of Cosmic Call points are classified as clusters that potentially contain anomalous signals of interest.
\item For each cluster containing anomalous signals we record the corresponding input file properties and the time of the event.
\end{enumerate}
This process is then repeated until all the input 2258 files are processed, and we have a large catalogue of potentially anomalous signals in the time-frequency domain.  In order to identify signals worth further investigation we then repeat the UMAP process using all the images identified as potentially containing an anomaly.  These then also get clustered and we finally inspect examples of each cluster by eye.

In more detail, for part~\ref{part3},
we  use ResNet50 which was first presented by \cite{heresnet}.
ResNet is efficient and fast, which has an accuracy of 95.9\% in identifing FRB in our previous work \citep{yangxuan}.

\begin{figure*}
	\includegraphics[width=0.98\textwidth]{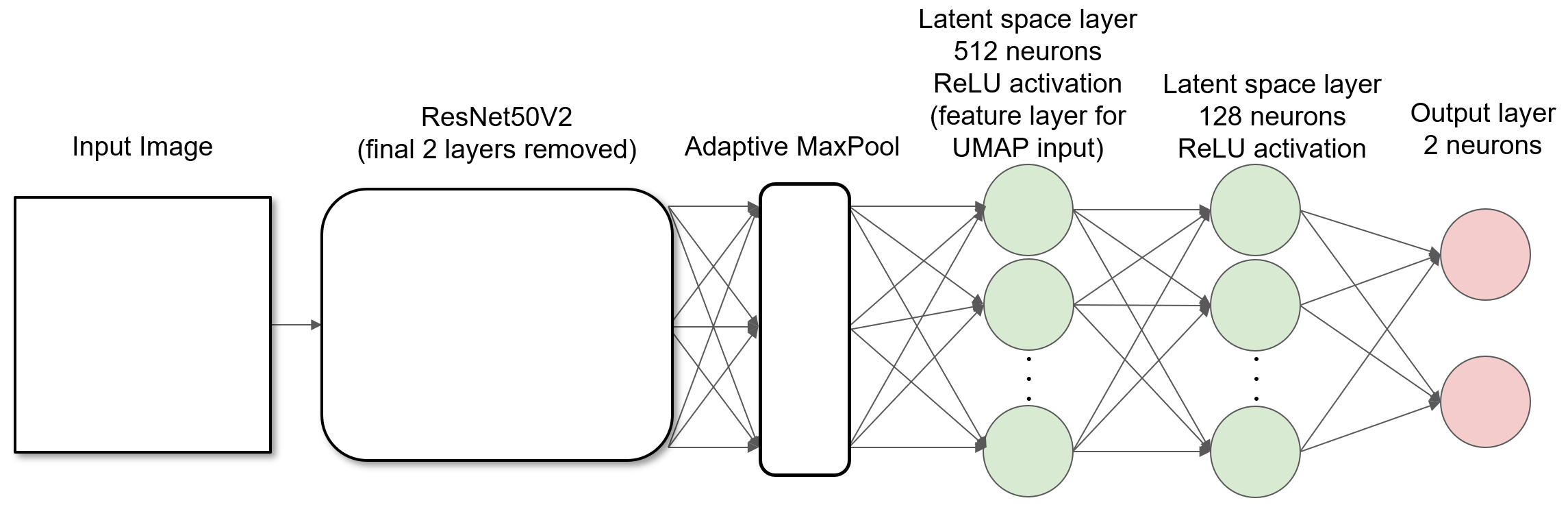}
    \caption{    
    The input $256\times256$ image serves as the initial input to the pre-trained ResNet50V2 model. The last two layers of the pre-trained network have been replaced with an adaptive max pooling layer. This is then followed by two fully connected layers, consisting of 512 and 128 neurons, respectively. Activation of these layers is achieved through the ReLU activation function. In this work, the output layer utilized is actually the 512 neuron layer.}
    \label{figure:resnet}
\end{figure*}

We restructured the ResNet as illustrated in Fig.~\ref{figure:resnet}. The final two layers of the original ResNet were an average pooling layer combined with a fully connected output layer used for classifying images. Since we do not need the classification function, we eliminated the final two layers and added an adaptive max pooling layer, along with two fully connected layers containing 512 and 128 neurons, respectively, activated by rectified linear units (ReLU). The model used parameters identical to those in our previous work for a similar task of extracting features from astronomical data~\citep{yang_umap}. After the ResNet model modification, the input image containing $256\times256 = 65,536$ pixels will be compressed and extracted features by the ResNet, and transformed to a 512D feature vector.

For part~\ref{part4}, we made use of a UMAP algorithm implemented in Python~\footnote{\url{umap-learn.readthedocs.io}}. To optimize the clustering results, we tuned two UMAP parameters: \emph{\sc n\_neighbours}, \emph{\sc min\_dist} and one hyper parameter \emph{\sc n\_samples}.
\emph{\sc n\_neighbours} represents the size of the local neighbourhood and affects the balance between local and global structure in the data. Larger values of \emph{\sc n\_neighbours} will make the UMAP focus on a broader structure of the data, but may lose more detailed information. 
\emph{\sc Min\_dist} controls the minimum distance between points in the low dimension and helps to preserve the broad topological structure. 
\emph{\sc n\_samples} means how many samples we feed into UMAP each time. The chosen values for the number of samples in each batch affect the result. Too small \emph{\sc n\_samples} may cause underfitting and lower the computation efficiency. Too large \emph{\sc n\_samples} may exceed the computation memory or cover up the outlier signals.
We performed a grid search for these parameters, \emph{\sc n\_neighbours} range from 2 to 1000, \emph{\sc min\_dist} range from 0.01 to 0.99, \emph{\sc n\_samples} range from 500 to 50,000. Finally \emph{\sc n\_neighbours=10}, \emph{\sc min\_dist=0.01} and \emph{\sc n\_samples=6,000} are determined for the P269-2001JANT dataset.

In part~\ref{part5}, instead of identifying regions by eye, we made use of ``spectral clustering'', which is a widely adopted algorithm to identify specific clusters of data points. This normally gives better results compared to traditional approaches such as k-means. The methodology behind spectral clustering lies in utilizing the eigenvectors of a similarity matrix to effectively partition the data into clusters. In this algorithm, the data points assume the role of nodes within a graph. It transforms the clustering task into a graph partitioning problem. 

The processing steps described above were implemented using multiple python scripts on the high-performance computer~(HPC) infrastructure provided by CSIRO. 

\subsection{Model evaluation}
We use the silhouette coefficient as a metric to assess the degree of cluster coherence~\citep{silhouettes}, which is defined as:
\begin{equation}
\begin{split}
    {\rm S(i)}=\frac{b(i)-a(i)}{max\left\{a(i),b(i) \right\}},
	\label{eq:silhouette}
\end{split}
\end{equation}
where $\rm S(i)$ is the silhouette coefficient value of point $i$, $a(i)$ is the mean distance between 
i and all other data points in the same cluster, $b(i)$ is mean distance of i to all points in the most neighboring cluster. Silhouette coefficient value ranges from $-$1 to 1. Values closer to 1 indicate well-defined clusters, while a coefficient greater than 0 indicates a satisfactory outcome.  

There are also two commonly used parameters in evaluating the performance of classification algorithms, precision rate and recall rate:
\begin{equation}
\begin{split}
    {\rm Recall}=\frac{T_{\rm p}}{T_{\rm p}+F_{\rm n}},
	\label{eq:recall}
\end{split}
\end{equation}
\begin{equation}
\begin{split}
    {\rm Precision}=\frac{T_{\rm p}}{T_{\rm p}+F_{\rm p}},
	\label{eq:precision}
\end{split}
\end{equation}
where $T_{\rm p}$ is true positive, indicating the number of anomalies correctly retrieved by the model in the anomaly clusters, $F_{\rm n}$ is false negative, indicating the number of repeaters incorrectly retrieved by the model in the non-detection clusters, and $F_{\rm p}$ is false positive, indicating the number of non-detection signals incorrectly retrieved by the model in the anomaly clusters.

These two parameters can be combined to get $F_\beta$ score~\citep{fscore}, which is defined as:
\begin{equation}
\begin{split}
    {\rm F_\beta}=(1+\beta^2)\cdot \frac{\rm Precision \cdot \rm Recall}{(\beta^2 \cdot \rm Precision)+\rm Recall},
	\label{eq:fscore}
\end{split}
\end{equation}
where $\beta$ is a weighting factor indicating that recall is considered $\beta$ times as important as precision. In our research, real anomalies are expected to be rare in the observational data. Therefore, we choose $\beta=2$~($F_2$ score) for the propose of finding more anomalies. The $F_2$ score means recall rate is two times more important than precision rate when evaluating our pipeline. 

In order to optimize the clustering results, a grid search was conducted, exploring a range of cluster numbers from 2 to 10. Throughout this search, the effectiveness of the spectral clustering was evaluated by considering the silhouette coefficient values and $F_2$ score. The silhouette coefficient is acceptable when larger than 0. We finally obtain the value 0.440 as an optimal value when we optimized the parameter \emph{\sc n\_cluster}, and \emph{\sc n\_cluster=8} performs the highest average silhouette coefficient value, which equals 0.440. The cluster number \emph{\sc n\_cluster=8} was finally chosen.

\section{Results and Discussion}
\label{sec:result}

\subsection{Rediscovery of the Lorimer burst}

\begin{figure*}
	\includegraphics[width=0.98\textwidth]{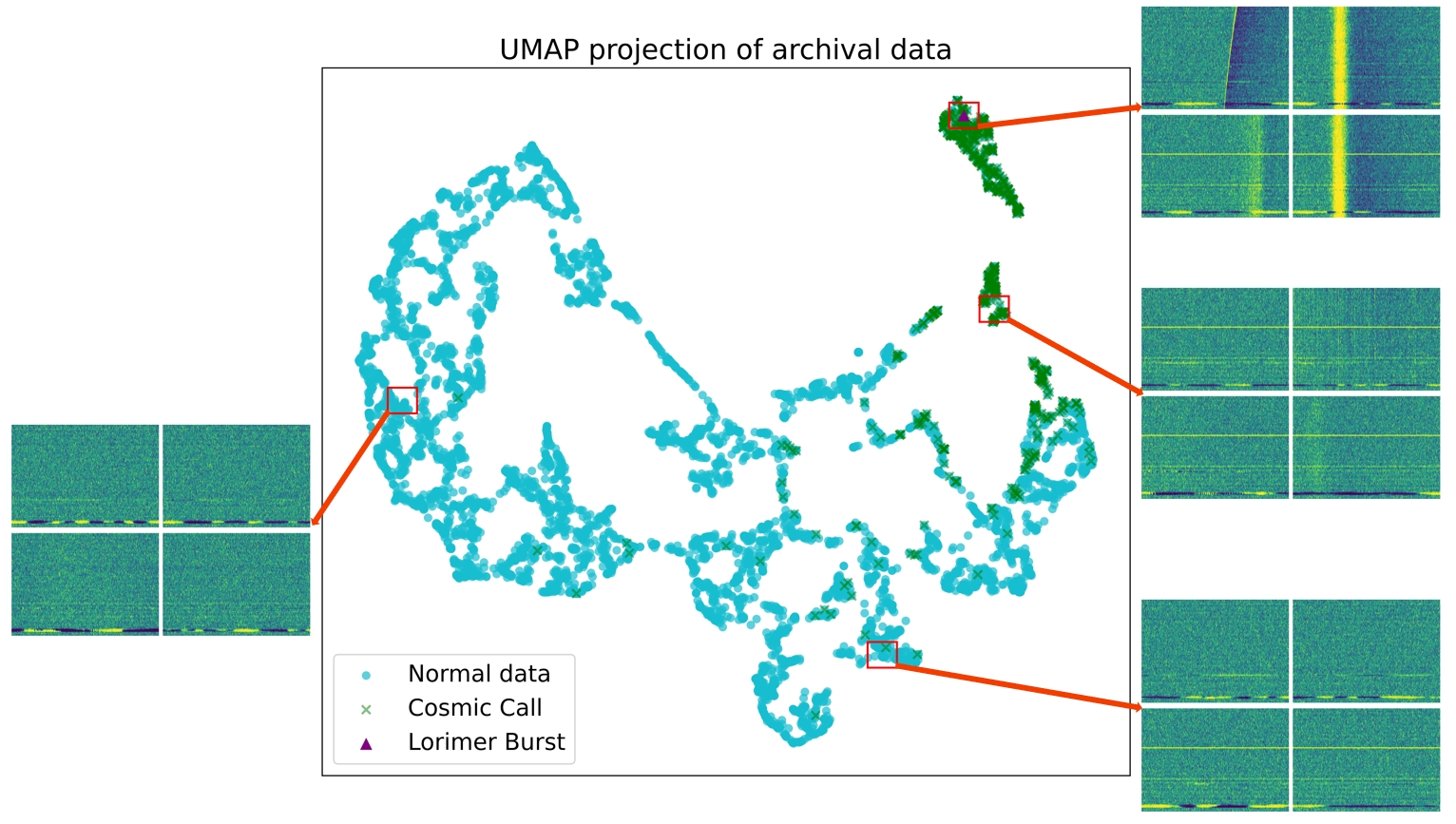}
    \caption{The UMAP embedding of 3 data files and cosmic call images. The pink triangle represents the Lorimer burst, and each cyan dot represents an image containing 4.096 seconds of observation. The four red squares show some representative images of the adjacent areas.}
    \label{figure:lorimer_cosmic_show}
\end{figure*}

In order to test our algorithm, we selected the data file containing the Lorimer burst along with data files corresponding to other beams in the same observation. We then processed these data in the same manner as described above.  Our UMAP result is shown in Fig.~\ref{figure:lorimer_cosmic_show}. The input images containing the Cosmic Call signals are shown as a green cross.  These generally cluster in the top-right region of the Figure.  
As the injected Cosmic Call signals become weaker then they form clusters that are slightly harder to distinguish from the normal data. Very roughly, the Figure divides into a few regions. On the left, the majority of the images contain no apparent signal, but can contain some persistent radio frequency interference~(RFI).  The bottom right region by eye looks very similar and no clear anomalous signals are seen. The strongest signals are in the upper right part of the Figure. We note that the second FRB discovered in this data set~\citep{songboburst2019} is undetectable by eye in the raw data stream and clusters with the ``Normal data'' in these UMAP results. 

\begin{figure*}
	\includegraphics[width=0.98\textwidth]{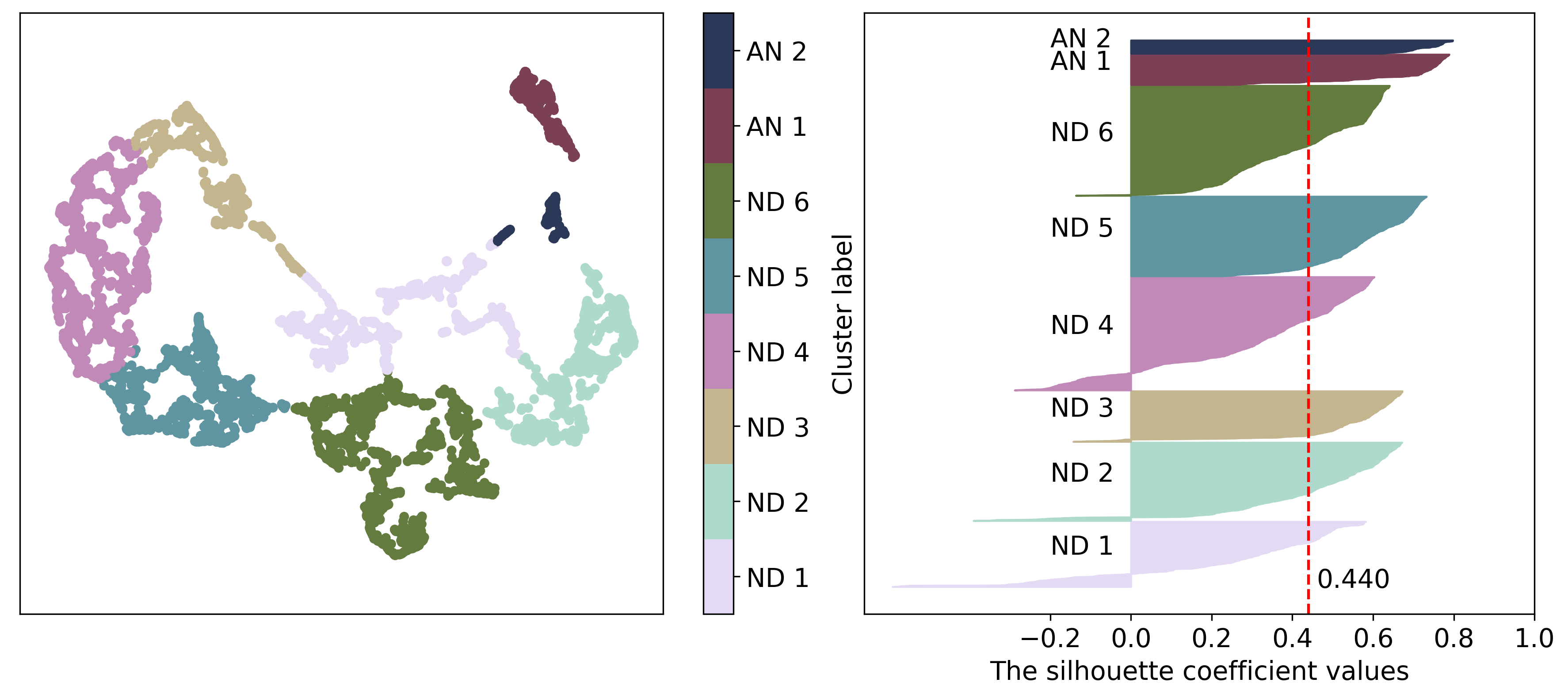}
    \caption{The left panel is the spectral clustering result of UMAP embedding in Fig~\ref{figure:lorimer_cosmic_show}. The label ND means non-detection cluster and AN means anomaly cluster. The right-hand panel corresponds to silhouette coefficient values. The red dotted line in the right panel represents the average silhouette coefficient value of the clusters.}
    \label{figure:lorimer_cosmic_spcluster}
\end{figure*}

With the help of spectral clustering algorithm, the embedding data points are separated into eight clusters shown in Fig.~\ref{figure:lorimer_cosmic_spcluster} left-hand panel.
The Cosmic Call points serve as tracers in each cluster, the density of Cosmic Call points indicates the probability of a cluster containing true anomalous signals. The clusters which contain more than 50\% of Cosmic Call points are classified as anomaly clusters (AN1 and AN2), 
other clusters are indicated as ND1 to ND6 where ND stands for non-detection of anomalous signals. Considering the total image quantity, this leads to a reasonable true positive and false positive rate.  Table.~\ref{table:lorimerburst} shows the number of candidates, how many Cosmic Call signals are within the cluster, and how many images are classified as being not an anomaly (non-detection). The data points in AN1 and AN2 were classified as anomaly candidates and further checked by eye. The Lorimer burst is easily detected in AN1. The AN1 cluster also contains some strong RFI. The AN2 contains some weaker signals which are confirmed as narrowband RFI.
In the right-hand panel of Fig.~\ref{figure:lorimer_cosmic_spcluster}, the silhouette coefficient values of the clusters are presented. The red dotted line represents the average silhouette coefficient.

\begin{table*}

\caption{The information of the spectral clustering result of UMAP embedding. The eight different clusters are identified in the searching pipeline.
There are two anomaly clusters, the data points in anomaly clusters are classified as anomaly candidates.\label{table:lorimerburst}}
\renewcommand\arraystretch{1.5}    
\begin{center}
\begin{tabular}{c|c|c|c|c}
\hline
\multicolumn{1}{c|}{ }  &     Total number        &     Cosmic Call number  	 &     	Anomaly candidate number  & Non-detection number        \\\hline

Non-detection 1    &   826  &   43   &   0     & 783     \\ 
Non-detection 2    &   992  &   125   &   0     & 867     \\
Non-detection 3    &   641  &   0   &   0     & 641     \\
Non-detection 4    &   1436  &   1   &   0     & 1435     \\
Non-detection 5    &   1009  &   4   &    0    & 1005   \\
Non-detection 6    &   1393  &   18   &    0   & 1375    \\ \hline 
Anomaly 1    &   386  &   366   &    20   & 0    \\
Anomaly 2    &   169  &   142   &    27   & 0    \\\hline 
\end{tabular}
\end{center}

\end{table*}

\begin{figure}
	\includegraphics[width=0.48\textwidth]{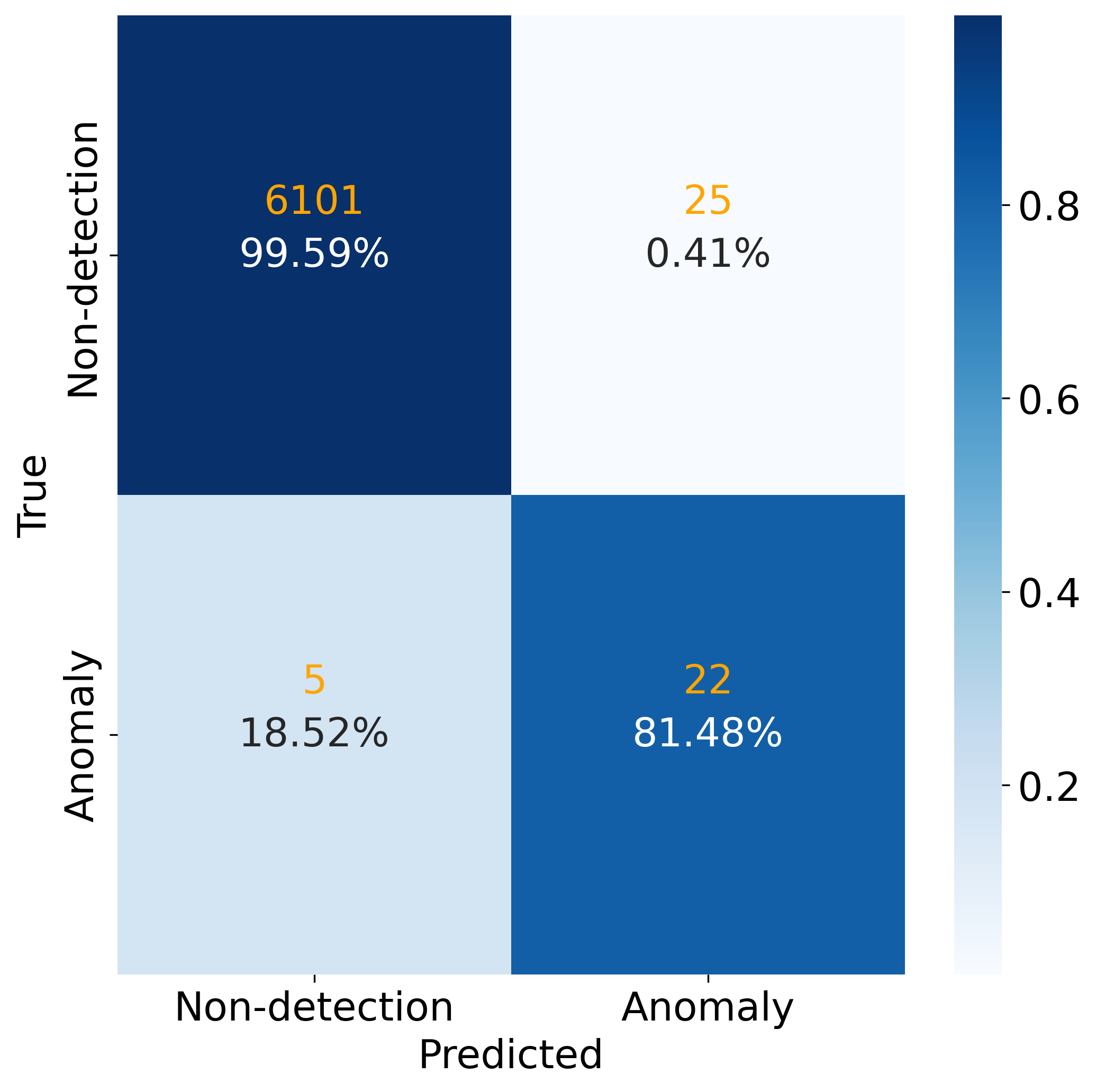}
    \caption{Confusion matrix of the UMAP searching result corresponding to Table \ref{table:lorimerburst}.}
    \label{figure:confusion_matrix}
\end{figure}

To obtain the confusion matrix and $F_2$ score of the searching pipeline, we manually classified all 6,153 images of observation data and found 27 anomalous images. Fig.~\ref{figure:confusion_matrix} shows the confusion matrix of the UMAP searching result corresponding to Table \ref{table:lorimerburst}.
We obtain $F_2=0.7097$.

The successful re-detection of the Lorimer burst provides confidence that our UMAP pipeline has the ability to find signals of potential interest. We therefore applied the pipeline to the entire P269-2001JANT data set.

\begin{figure}
	\includegraphics[width=0.48\textwidth]{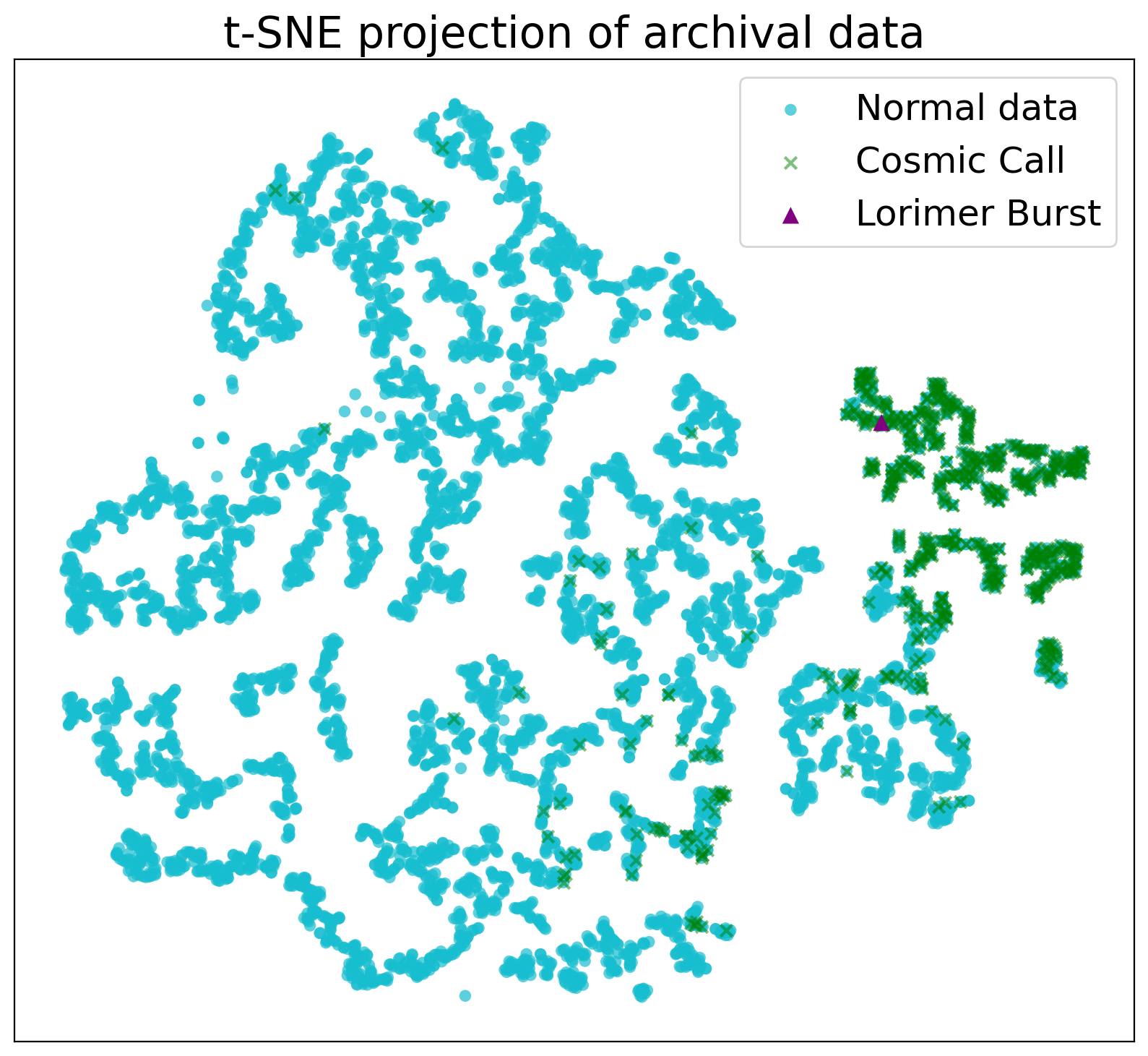}
    \caption{The t-SNE embedding of 3 data files and cosmic call images. The pink triangle represents the Lorimer burst, and each cyan dot represents an image containing 4.096 seconds of observation.}
    \label{figure:tsne}
\end{figure}

\subsection{Comparison with other searching methods}

t-SNE is also a popular nonlinear dimensionality reduction technique. UMAP and t-SNE can produce similar results in single-cell data analysis~\citep{umap5}. For comparison, we process the same data as mentioned above with t-SNE and the result is shown in Fig.~\ref{figure:tsne}. Most of the green crosses in t-SNE embedding are also clustered in the upper part of the right-hand region. A small number of blue points which contain strong anomalous signals are distributed in  this region. A few, weaker, Cosmic Call signals are detected in lower middle of Fig.~\ref{figure:tsne}. In general, the t-SNE embedding in Fig.~\ref{figure:tsne} looks similar to UMAP embedding in Fig.~\ref{figure:lorimer_cosmic_show}. However, the t-SNE results have lost more of the global structure of the data than UMAP. 
The global and local here are defined by parameter \emph{\sc n\_neighbours}, a point and its \emph{\sc n\_neighbours} number of closest neighbours are considered as the local scale. 
The missing global structure makes spectral clustering algorithm unable to classify clusters from the embedding. In addition, t-SNE takes more than five times as long to process the data than the UMAP method.

We also processed the data using a single pulse searching pipeline. The data was first searched by \emph{\sc HEIMDALL}\footnote{\url{https://sourceforge.net/projects/heimdall-astro/}} using S/N $\geq7$, DM $\leq5000$ and width $\leq4.096s$. \emph{\sc HEIMDALL} found 25.6\% of injected Cosmic Call signals and 11.1\% of observed anomalous signals. The candidates were then plotted and classified by the deep neural networks~\emph{\sc FETCH}~\citep{FETCH}. \texttt{FETCH} rejected most of the candidates provided by \emph{\sc HEIMDALL}, and finally found 5.9\% of injected Cosmic Call signals and 11.1\% of observed anomalous signals. And this pipeline can find the weaker FRB~\footnote{The S/N and observed width of this burst are 11 and 24.3\,ms, respectively.} published by \cite{songboburst2019}. This algorithm has good performance in searching for FRBs, but missed most of the injected anomalies. 

\subsection{Anomalies in the data set}


After processing our entire data set, 27,308 images were identified as containing potential anomalies from the 2,320,167 raw data images.  We need to distinguish astronomical sources from terrestrial sources of interference.  With a single-dish radio telescope, this is non-trivial~\citep{blc1}.  However, the use of the multibeam receiver allows some ability to distinguish ground-based signals from astronomical signals.  In particular, any weak astronomical signal would only be seen in a single beam.  For very strong signals (such as the Lorimer burst) we would expect to detect the same signal in a few adjacent beams.  Any signal seen in a large number of beams or in widely separated beams can therefore be considered to be interference.
We automatically searched for candidates seen simultaneously from more than three beams and rejected these as being caused by RFI.  After this filter, the number of images decreased to 10,451. 

Finally, a visual inspection was conducted on all of these images.  The UMAP process did often miss weak events in some beams and therefore we carried out a visual inspection on these remaining candidates.  We found that the majority were weak events in all 13 beams and hence rejected these manually, leaving us with 202 potential events. Most of these were detected in only one beam of the multibeam system. In no case have we observed signals that appear in multiple beams but not in adjacent ones. An online Table in~\url{https://astroyx.github.io/anomaly/} lists the properties of these anomalies. 
 
\begin{figure*}
	\includegraphics[width=0.98\textwidth]{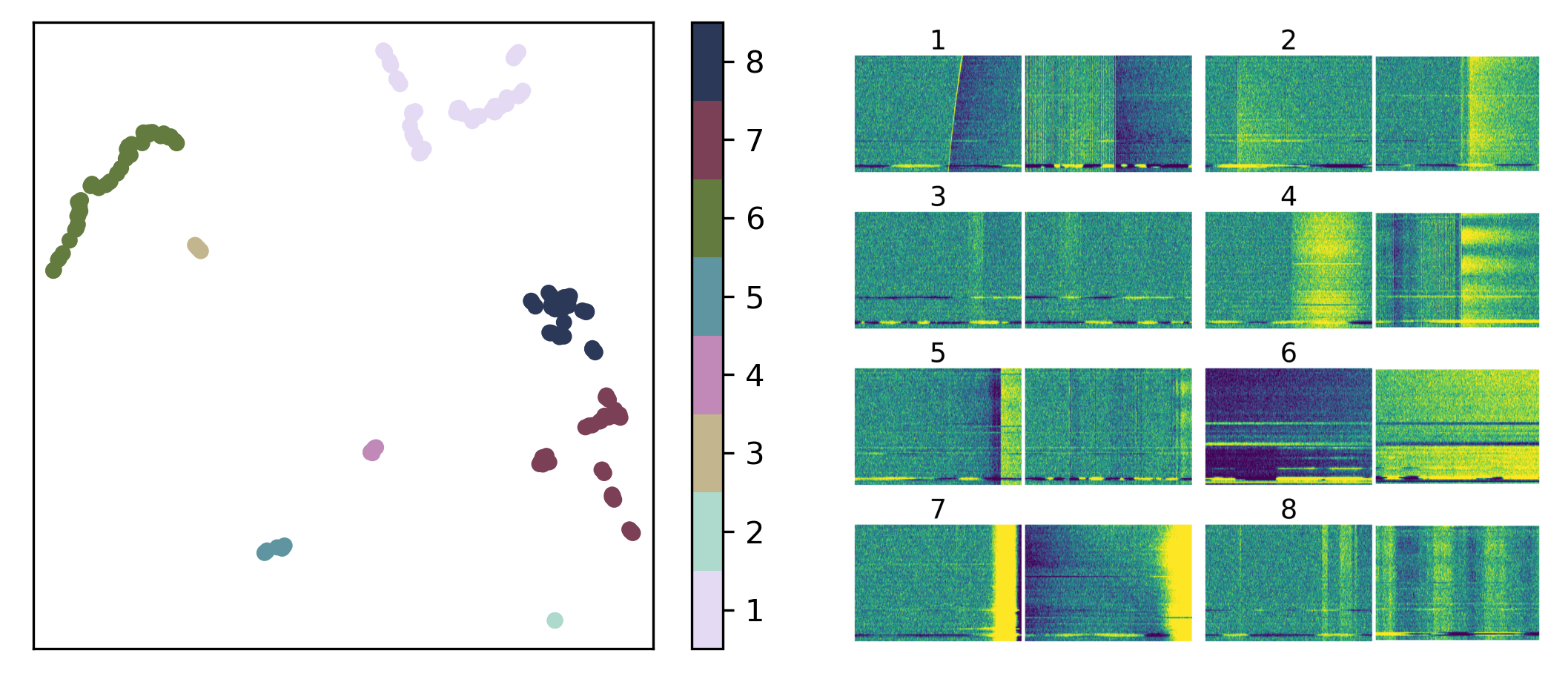}
    \caption{ The left-hand panel shows the UMAP embedding of 202 anomalous events and Lorimer burst, the clusters are marked with different colours named from one to eight. The right-hand panel shows two representative figures of each cluster.}
    \label{figure:anomaly_spcluster}
\end{figure*}

In order to search for commonalities between these 202 events, we show the UMAP embedding of these anomalous events in the left-hand panel of
Fig.~\ref{figure:anomaly_spcluster}. The embedding was also divided into eight clusters using spectral clustering and marked with different colours. 
The right-hand panel of Fig.~\ref{figure:anomaly_spcluster} shows two representative figures of each cluster. We can find some morphological similarities inside each cluster. For example, cluster one contains images with bright signals where the digitizer saturated during the event and took time to recover, which includes the Lorimer burst. The images in cluster six are dominated by saturation effects occurring after bright events, although these events only happened in one beam of the multibeam system. In most cases, this is likely to be a strong RFI signal in the pointing direction of that particular beam (the beam patterns for a source in the sky are described by~\citealt{multibeam_snr}), but these sources are worthy of follow-up study and we discuss one such event below. The images corresponding to cluster eight show broadband signals each lasting around a second, but only detected in a single beam. On this scale, we see no evidence of any dispersion in these signals. Signals with these characteristics occur several times at various positions in the sky throughout a relatively short data span (the data span is from March to June in 2001) and persist for a few seconds on each occurrence. These events do not always occur in the same beam and the sky position varies. Such events could occur from distant lightning, 
 but we are currently unable to confirm that conjecture.
The other clusters are similar. 
The majority of the events are seen in a single beam, however, events in cluster seven have been detected in up to three adjacent beams. 



We currently have no easy means to determine which of these 202 events are worthy of follow-up (noting that for a future survey, we would apply these methods in close to real-time, allowing for a much quicker response time).  Instead, we have chosen two events for a more detailed analysis. 

\begin{figure*}
	\includegraphics[width=0.98\textwidth]{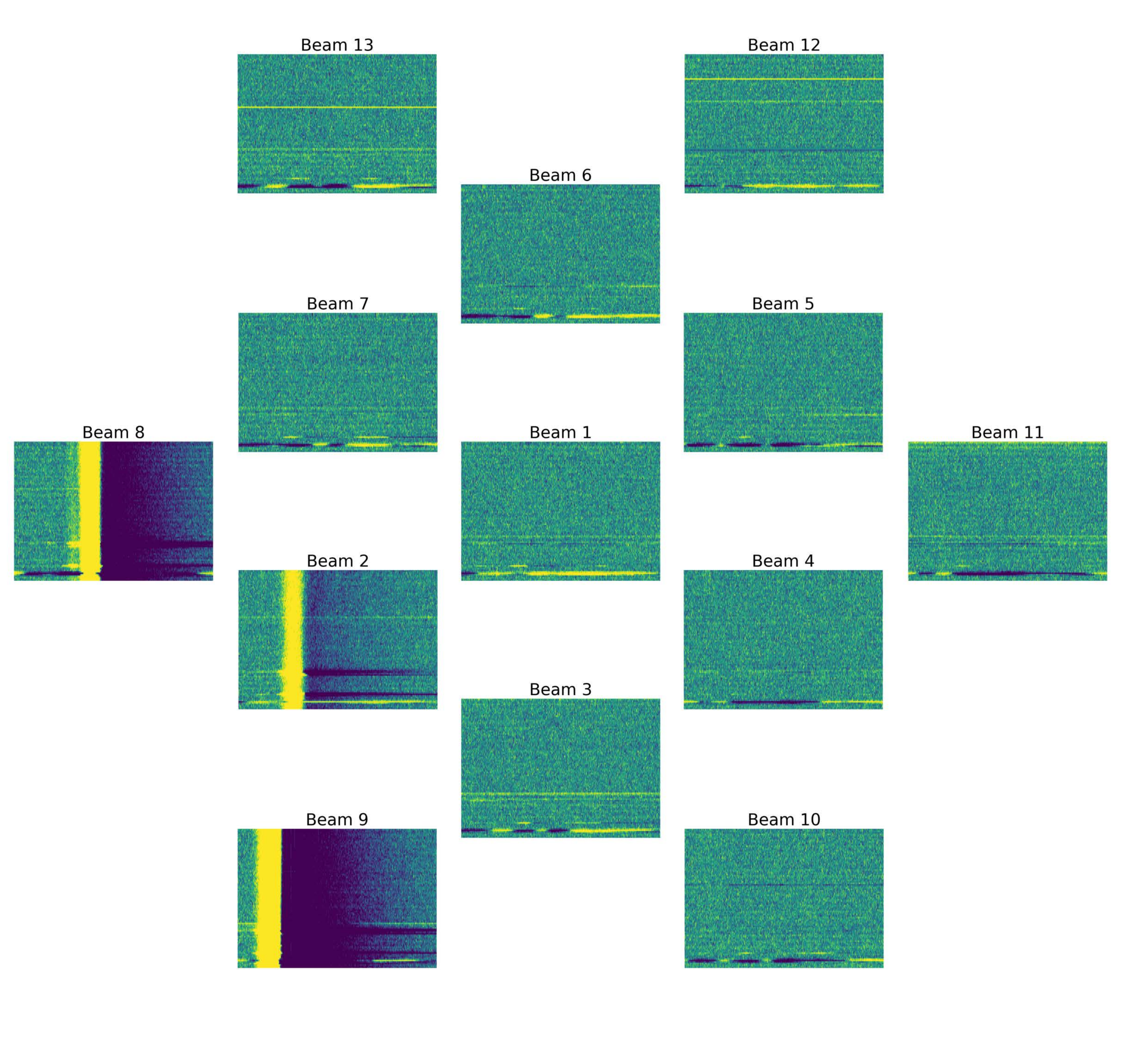}
    \caption{One of the anomalies from SMC019\_01521, starts at time sample 5758976, and ends at 5763072. Each figure represents a time span of 4.096 seconds along the x-axis, 1,516.5 to 1,231.5 MHz from bottom to top along the y-axis.}
    \label{figure:p269_ano_3}
\end{figure*}

In Fig.~\ref{figure:p269_ano_3}, we show one example where a bright event is seen in three adjacent beams.  In this Figure, we show the region around the event in all 13 beams of the multibeam receiver with the panels approximately representing the positions of the beams. The UMAP analysis is based on individual images presented to the algorithm.  It does not account for any differences in the time of an event during those images.  This is shown here where the signal is first seen in Beam 9, then less than a second later in Beam 2 and then in Beam 8. By setting the arrival time of Beam 9 as 0 sec, the arrival times of Beam 2 and Beam 8 are determined to be 0.618 sec and 1.007 sec, respectively. Because of this time difference, it is clear that this particular event is not an astronomical signal of interest and is representative of a satellite moving across the beams.



\begin{figure*}
	\includegraphics[width=0.98\textwidth]{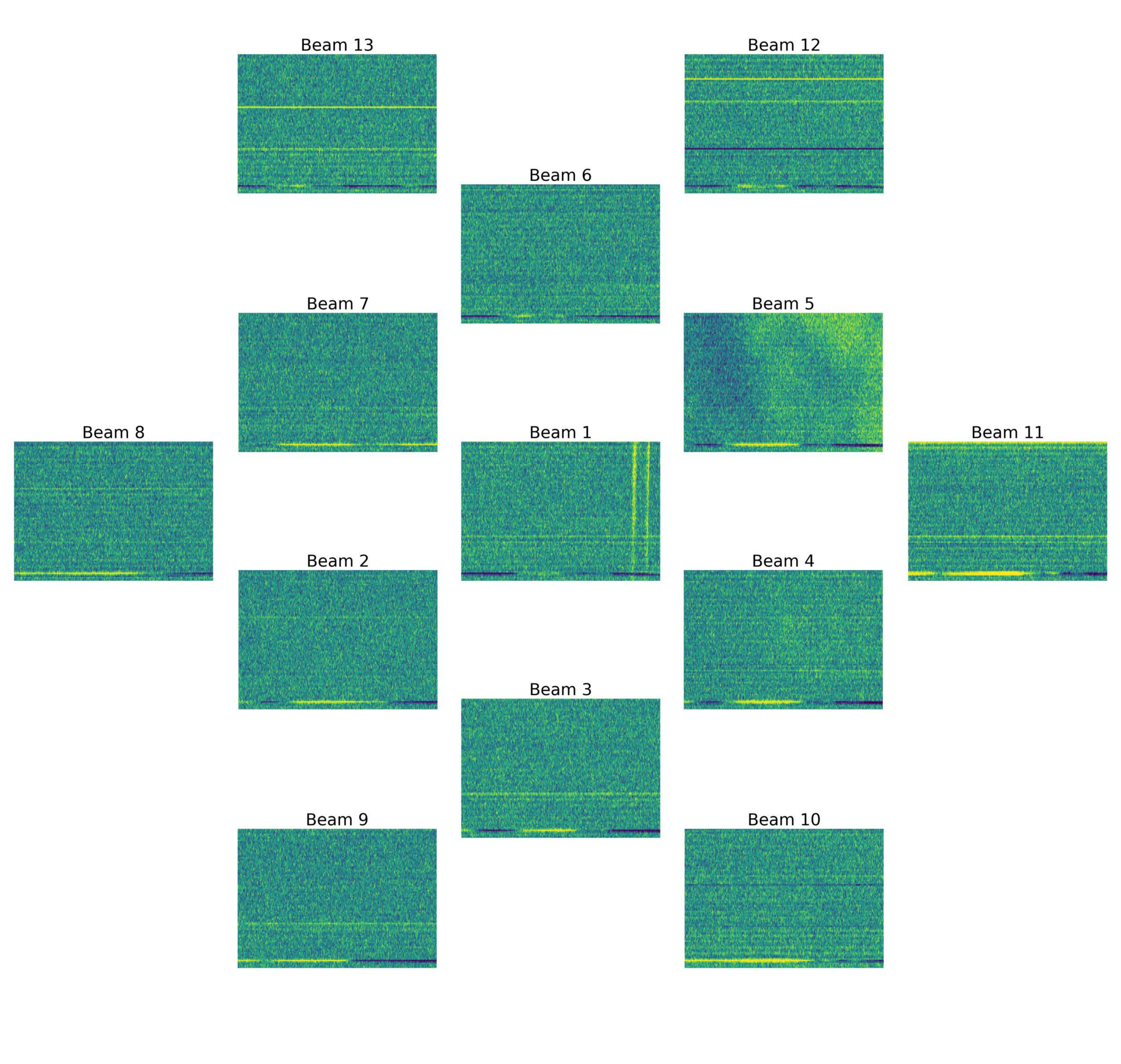}

    \caption{One of the anomalies from SMC017\_02351.sf, starts at time sample 159744, and ends at 163840. The beam 1 was pointing at PSR~J2048$-$1616. The anomaly detected by UMAP is in beam 5, and the other figures are the multibeam receiver's observation simultaneously. Each figure represents 1.024 seconds of observation along the x-axis, 1,516.5 to 1,231.5 MHz from bottom to top along the y-axis. The GIF version of the one-minute multibeam observation can be found at~\url{https://astroyx.github.io/anomaly/}, GIF column.}
    \label{figure:p269_ano_cloud}
\end{figure*}

One of our most interesting sources is shown in Fig.~\ref{figure:p269_ano_cloud}.  In this figure, we see a broadband signal that lasts a few seconds. However, through the 60-second observation, it repeats.  These particular set of observations correspond to a test sequence of the multibeam system, where each beam in turn was pointed to the pulsar, PSR~J2048$-$1616.  We note that this pulsar is bright and it is possible to detect single pulses from the pulsar (which have a DM of 11.46\,cm$^{-3}$pc and a width of 12ms). However, the pulsar is always in a different beam from the signal of interest here and our signal was detected in eight of the 60-second observations with the source being consistently located at right ascension 20:47:34.4 and declination $-$16:42:03~\footnote{The multibeam system covered the position eight times with different beams, the standard deviation of the eight pointing directions in right ascension is 0.6 arc-sec, and 9.0 arc-sec for declination. The sky coverage of each beam is $\sim0.23$~deg.}. 

We thus propose that these repeating signals are likely to originate from an extraterrestrial source.

\begin{figure}
	\includegraphics[width=0.47\textwidth]{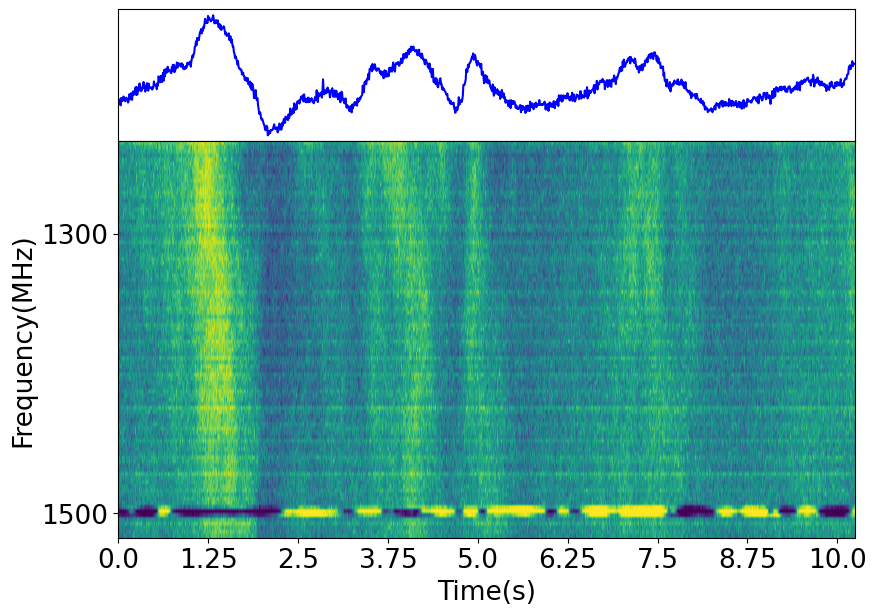}

    \caption{Pulse profiles of the anomaly from SMC017\_02351.sf, the bottom panel shows the time–frequency plane, while the top panel is the integrated pulse profile using an arbitrary flux density scale.}
    \label{figure:cloud_10sec}
\end{figure}

Images of these events are provided online at~\url{https://astroyx.github.io/anomaly/}.  
We applied two independent approaches to search for the DM of these bursts: (1) single-pulse searches using \emph{\sc PRESTO}, and (2) manual maximization of the burst SNR across a range of  DM values from 0 to 1000\,cm$^{-3}$pc. However, both methods failed to yield a significant DM detection.
Fig.~\ref{figure:cloud_10sec} shows examples of the dynamic spectrum for the detected signals, with the x-axis spanning 10.24 seconds and the y-axis covering the frequency range from  1231.5 to 1516.5\,MHz. The Milky Way's DM contribution on this sky coordinates~(galactic longitude 29$^\circ$.9384, galactic latitude -33$^\circ$.0078) estimated using YMW16 electron-density model is 40.78\,cm$^{-3}$pc~\citep{Yao2017}.
Notably, within this frequency range, even a DM$\sim$ 100 \,cm$^{-3}$pc would result in a time delay of only
$\sim0.1$ s from the highest frequency to lowest frequency, which is negligible compared to the duration of the detected bursts. Determining a valid DM estimate for this source is challenging at this frequency band.

We have applied a 15 arc-min cone search in the CSIRO data archive~\citep{george_archive} for this sky position and identified 11 observation files after P269-2001JANT.  We have searched these files by eye and have not identified any similar event. 
 
Within a 0.23\,deg radius, GAIA identified 2,524 sources and 877 of them have distance measurements from GSP-Phot Aeneas best library using BP/RP spectra. 
Due to poor localization and the lack of a reliable DM estimate, investigating the origin of this source remains challenging.
We propose long-term follow-up observations of this sky region using the ultra-wideband (UWL) receiver on the Parkes telescope. The UWL receiver covers a frequency range from 704 to 4032 MHz, which could enable better studies of the DM, pulse profile, and periodicity of the source if more bursts are detected. 
Furthermore, observations at even lower frequencies, which are expected to show larger dispersion time delays and provide good DM detection capability for second-duration bursts~\citep{mwa_white_dwarf}, as well as observations with array telescopes offering more precise localization~\citep{Wang2024}, are encouraged to explore the origin of this source.   

\section{Conclusion}
\label{sec:Dis}

The UMAP unsupervised machine learning pipeline has the potential to identify anomalous signals in high-time-resolution data sets. Applying this pipeline to the P269-2001JANT dataset, we rediscovered the Lorimer burst from the dataset~\citep{Lorimer:2007qn} and also identified a relatively large number of other events. A table containing the properties of these events is provided online. 
The majority of these events have durations on a timescale of seconds, which corresponds to several solar radii. 
We note that because these observations were taken over two decades ago it may be impossible to identify the cause of many of these signals and hence this method has primarily been designed for the next generation of surveys carried out with the Murriyang Parkes radio telescope. \cite{peryton} found radio bursts with similar spectra to FRBs, but they were confirmed as terrestrial origins because of 13 beams simultaneously detection. The multibeam information is crucial for identifying signals originating from the universe.

A cryogenic phased array receiver~(cryoPAF) is currently being completed and wll be installed on the telescope in 2024. This receiver will increase the number of beams from 13 to 72 and the bit depth of the backend instrument will increase from the 1-bit data used in this paper to 2-bits or 8-bits per sample.  Both of these improvements will enhance the UMAP algorithm.  The higher bit precision will allow the features in the data streams to be better characterised. 
The large number of beams and more flexible backend system of the cryoPAF will allow us to develop new algorithms to mitigate RFI~\citep{paf}, and we plan to develop the UMAP pipeline to explicitly incorporate the spatial information of the multiple beam signals. 

The data processed here were all available in a single archive.  This method of processing data will remain viable assuming that the raw data sets are archived.  However, the increasing data volumes from radio telescopes often means that searches for transient events are carried out in real-time and only data surrounding events of interest are stored.  Our pipeline can easily be converted into a ``real-time'' mode.  An almost identical data set for clustering would occur if instead of picking 2,051 subintegrations from each of the three data files, we instead selected 100 subintegrations from 72 input files. This would require a larger memory capacity, IO speed and a parallel computing accelerator 
to accommodate this real-time processing.

\section*{Acknowledgements}

This paper includes archived data obtained through the Parkes Pulsar Data Archive on the CSIRO Data Access Portal~(\url{https:// data.csiro.au/}). The computing resource is offered by Scientific Computing’s shared system in CSIRO. 
This work is partially supported by the National SKA Program of China (2022SKA0130100,2020SKA0120300), the National Natural Science Foundation of China (grant Nos. 12041306, 12273113,12233002,12003028,12321003), the CAS Project for Young Scientists in Basic Research (Grant No. YSBR-063), the International Partnership Program of Chinese Academy of Sciences for Grand Challenges (114332KYSB20210018), the National Key R\&D Program of China (2021YFA0718500), the ACAMAR Postdoctoral Fellow, China Postdoctoral Science Foundation (grant No. 2020M681758), and the Natural Science Foundation of Jiangsu Province (grant Nos. BK20210998).

\section*{DATA AVAILABILITY STATEMENTS}
The data used in this work is available from \url{https://data.csiro.au/}. Anomaly figures are available from \url{https://astroyx.github.io/anomaly/}.





\bibliographystyle{mnras}
\bibliography{mnras} 

\bsp	
\label{lastpage}
\end{document}